\documentclass[twoside]{article}
\usepackage[latin1]{inputenc}
\usepackage{t1enc}
\usepackage{a4}
\usepackage{tabularx}
\usepackage{epsfig}

\def\bref{\vspace{4pt}\noindent\hangindent=10mm}

\newcommand{\mym}{{\rm \mu m}}
\newcommand{\kpc}{{\rm kpc}}
\newcommand{\pc}{{\rm pc}}
\newcommand{\gyr}{{\rm Gyr}}
\newcommand{\myr}{{\rm Myr}}

\newcommand{\yr}{{\rm yr}}
\newcommand{\kmskpc}{{\rm km}\,{\rm s}^{-1}\,{\rm kpc}^{-1}}
\newcommand{\kms}{{\rm km}\,{\rm s}^{-1}}
\newcommand{\mhz}{{\rm MHz}}
\newcommand{\etal}{{\em et al}}
\newcommand{\xx}[1]{}

\begin{document}

\setcounter{figure}{0}
\setcounter{section}{0}
\setcounter{equation}{0}

\begin{center}
{\Large\bf
Gas Streams and Spiral Structure in the\\[0.2cm]
Milky Way}\\[0.7cm]

Peter Englmaier \\[0.17cm]
MPI f\"ur extraterrestrische Physik\\
Postfach 1603, D-85748 Garching \\
ppe@mpe.mpg.de
\end{center}

\vspace{0.5cm}

\begin{abstract}
\noindent{\it
  The observed gas dynamics in the Milky Way can only be explained by a
bar in the galactic center. Such a bar is directly visible in the
near-IR maps of the bulge, where it causes a distinctive asymmetric 
light distribution pattern. Another large-scale structure is the
grand-design 4-arm spiral pattern, most clearly observed in the
spatial distribution of molecular gas and HII-regions. Since this spiral 
arm pattern is regular, it must have existed for at least a few rotations.
Traced by molecular clouds, the spiral arms appear strongest in the
so-called molecular ring between 4 and 7 $\kpc$. 

In order to model the observed gas flow structure, we constructed a
model for the stellar mass distribution. For the inner 5 $\kpc$ we used
the 3D deprojected near-IR light distribution, as observed by the
COBE/DIRBE experiment, and added an analytical disk model outside the
box as well as a halo model.  With this frozen mass distribution, we
computed the stationary gas flow for various deprojection parameters
and pattern speeds.  For all reasonable parameter choices, we obtain a
4-armed spiral pattern, which can be matched to the observed spiral
arms. The model spiral arms, are caused by the forcing of the bar, and
two additional mass concentrations at about 4 kpc's on the minor axis
of the bar, which are likely to be incorrectly deprojected stellar
spiral arms.

In the bar region, our model can explain the non-circular motion
visible in the terminal velocity curve as well as some part of the
forbidden velocities. Inside the corotation, we also find 4 spiral
arms, the nearest arm corresponds to the 3-kpc-arm, although only
qualitatively. The missing southern 3-kpc-arm at the far end of the
galaxy is explained by running parallel to another arm.  Close to the
center, we find gas on circular orbits forming a disk. Such a disk has
been observed in emission of the CS molecule, however only part of the
disk appears to be occupied by dense enough gas to be traced by CS.
}
\end{abstract}

\section{Introduction}

The Milky Way is our local laboratory of star formation and ISM
physics.  Some observations even have only been possible in the Milky
Way, e.g.  turbulent motion and magnetic field in ISM clouds, or
search for dark matter candidates.  Also, many observable quantities
are directly linked to the history and structure of the Milky Way,
e.g. metalicity gradients (Friedli 1999), and distribution of OH/IR
stars (see below). A better understanding of the detailed mass
distribution and gas flow will therefore benefit many other studies.

Morphology and type of the Milky Way are hard to recognize because of
dust obscuration and the location of the sun within the disk.  Spiral
arm tangents can be identified in optical and radio brightness
distributions (Sofue 1973), as well as in molecular gas (e.g., Dame 
\etal. 1987), HII regions (Georgelin \& Georgelin 1976), and other
tracers; see Vall\'ee (1995) for a complete list.  The 3-kpc-arm, is
unusual in this respect: it is very bright in CO, but not traced by
HII regions. It is therefore possible that it does not form stars at the
present time. Maybe this is because it has an enhanced turbulent velocity or
a higher differential shear (Rohlfs \& Kreitschmann 1987).  The 3-kpc
arm and other peculiar gas dynamics in the galactic center are evidence for a
bar in the bulge region. The other spiral arms outside the bar region
within the so-called molecular ring appear to be on almost circular
orbits since they do not show large non-circular motion in front of
the galactic center. A comparison of many publications indicates that
the Milky Way most likely has 4 spiral arms with a pitch angle of
{12$^\circ$} (Vall\'ee 1995).

Another important aspect is, that the true rotation curve of the Milky
Way can be measured more precisely by a combination of a stellar mass
model and a hydrodynamical model. When the rotation curve is determined
from the gas dynamics alone, an axisymmetric model yields an uncommon
sharp peak at $\sim0.5\,\kpc$ (Clemens 1985). An axisymmetric model
of the stellar mass distribution, however, implies a mostly flat
rotation curve there (Kent 1992). A bar naturally solves this issue by
explaining the sharp peak with non-circular motion caused by orbits
elongated along the bar (Binney \etal. 1991). It also explains the
observed nuclear disk of molecular gas inside $\sim200\,\pc$.

Finally, having a better model for the mass distribution and rotation
curve, one also obtains better constraints on the amount of dark
matter in the solar neighborhood.

Since a short review cannot give a complete coverage of all Milky Way
models in the literature, we refer the reader to the following papers
for similar and alternative models: Lin, Yuan \& Shu (1969),
Mulder \& Liem (1986), Amaral \& L\'epine (1997), 
Wada \etal. (1994), Weiner \& Sellwood (1999), and Fux (1999a).
For reviews about spiral structure see: Wielen (1974), Toomre (1977),
Binney \& Tremaine (1987), and Bertin \& Lin (1996).

\section{Using observations to yield a detailed mass model}

An early axisymmetric mass model for the inner Galaxy was constructed
by Kent, Dame \& Fazio (1991), by fitting parametric models to the
photometric near-IR maps at $2.4\,\mym$ (K-Band) which where
obtained by the Spacelab Infrared Telescope (IRT) and have a
resolution of $1^\circ$. By assuming a constant mass-to-light ratio
for each component Kent (1992) found various possible combinations of
bulge, disk, and halo components to fit the observed mass distribution
and kinematics, as well as the gaseous rotation curve outside the
bulge region.  The bulge region was excluded from the fit, because the
gaseous rotation curve from Clemens (1985) shows signs of non-circular
motion due to the presence of a bar. Since the dark halo is only
observable through its gravitational interaction, Kent used the dark halo
component to compensate for the mismatch between gaseous rotation
curve and the rotation curve implied by the disk and the mass
distribution in the inner galaxy. He also defined a maximum disk
model, which minimizes the amount of dark matter required by
maximizing the mass-to-light ratio for the disk (he found $M/L =
1.3$). This model is particular interesting, because it gives a lower
limit for the amount of dark matter in the solar neighborhood. One
reason why there is large uncertainty about the relative contribution
of dark halo and disk, is that not enough constraints about the
vertical distribution of mass in the galaxy are known (Dehnen \&
Binney 1998).

A major improvement over the model of Kent was made possible by the
near-IR maps obtained with the DIRBE experiment on board of the COBE
satellite.  By using a foreground dust screen model, Dwek 
\etal. (1995) found, that the bulge light is best fitted by a triaxial
light distribution. This bar appears to be elongated with axis ratios
$1:0.33:0.22$, the nearer end at positive longitudes and inclined by
about $20^\circ$. Later, an improved dust correction and improved
parametric model was obtained by Freudenreich (1998).

Another line of models comes from a non-parametric deprojection method
introduced by Binney \& Gerhard (1996). No parametric model
distribution has to be prescribed, but an initial model has to be
given. The method can also be applied to a part of the galaxy, while
the model outside that part stays fixed as specified by the initial
parametric model.  Binney, Gerhard, \& Spergel (1997) applied the
method to the DIRBE data and found, that the result is robust against
changes in the initial model and can reliably recover the 3D light
distribution in the bulge. They also found additional light
concentrations on the minor axis of the bar, which they attribute to
spiral arm heads in that area. While the deprojection method is not
able to recover spiral arm structure, tests show that spiral arms
could produce clumps on the minor axis like the ones found in the
DIRBE data. The only free parameters in this model are the
mass-to-light ratio and the orientation of the bar.  See Gerhard
(1996, 1999) for a comparison of these models and further evidence for
the bar.

For further modeling, we expanded the mass model from Binney, Gerhard,
\& Spergel (1997) into a series of spherical harmonics to calculate
the gravitational potential.  The radial density profile in the
galactic center follows a power law $\rho\propto r^{-1.85}$ (Becklin \&
Neugebauer 1968) and this peak in the density is not reproduced by the
deprojection because of finite resolution and smoothing effects. We
therefore corrected the zeroth order spherical harmonic (monopole) to
reflect the power law. 

The resulting rotation curve falls off beyond $4.5\,\kpc$, because
no dark halo has been included yet. By construction, the model should
reflect the contribution of all luminous mass in the galaxy without
any assumptions about how bulge and disk are build together assuming
the same M/L. For some models, we changed the monopole to allow for a
flat rotation curve. 

\section{Terminal rotation curve}

The rotation curve of the Milky Way is not directly observable.  In
external galaxies, the rotation curve can be obtained by an
observation of Doppler shift along a slit across the center. Using
parametric models for disk, bulge and halo, the rotation curve can
then be decomposed. The parametric model for disk and bulge must
follow the observed surface brightness, while the dark matter halo is
only constrained by the fit to the rotation curve. Where density
wave theory is applicable, i.e. for tightly wound spirals, further
constraints are available for the dark matter contribution
(Fuchs, M\"ollenhoff \& Heidt 1998).

In the Milky Way, the rotation curve can only be inferred indirectly
from the terminal velocity, i.e. the maximum observed radial velocity
within the galactic plane at a given longitude. Historically, the
terminal rotation curve, has been used to infer the radial mass
distribution by assuming circular rotation. Inside the solar circle,
the terminal velocity is equal to the circular rotation curve minus
the motion of the local standard of rest (LSR).  The motion of the LSR
can be inferred from the streaming of stars measured by Hipparcos
(Feast \& Whitelook 1997), or the proper motion of Sgr~A$^*$ (Backer
\& Stramek 1999). Both methods agree reasonable well, and no $m=1$
mode seems to be present in the center of the galaxy.

\begin{figure}[!t]
\begin{center}
\epsfxsize=11cm \epsfbox{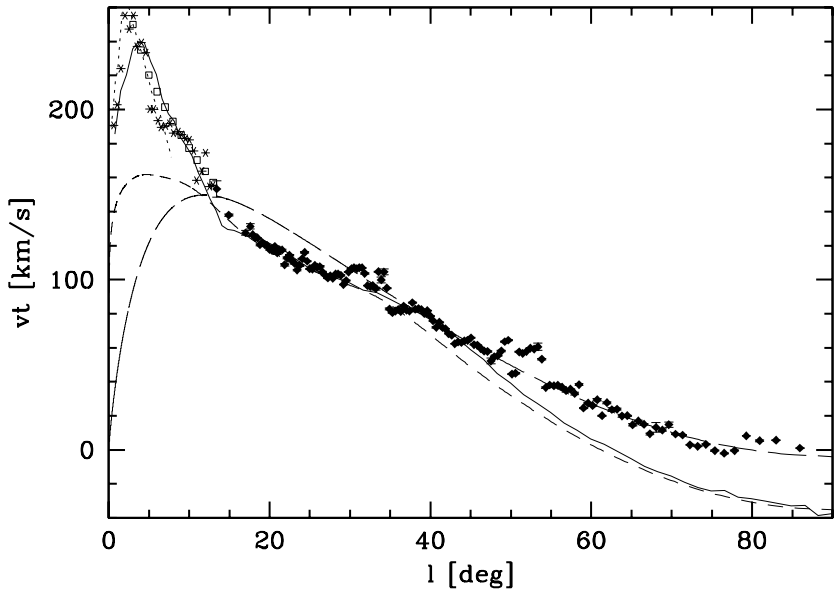}
\vskip -7mm
\end{center}
\caption{Terminal curve for Kent (1992, long-dashed line),
axisymmetric BGS model without halo (short dashed line), and gas
dynamics in full BGS model without halo (solid line and dotted line with
higher resolution).  }
\label{termcmp}
\end{figure}

The high peak in the terminal curve, gas on apparently forbidden
velocities, and the large radial velocity of the 3-kpc-arm indicate
non-circular motion in the inner galaxy.  This led to the idea, that
our galaxy is actually barred and the presence of a bar implies
non-circular motion which cannot be corrected for without
modeling (Gerhard \& Vietri 1986). 
Fortunately, the non-axisymmetric light distribution of the
bar can be extracted from photometric imaging due to perspective
effects.  These effects are only strong enough if the diameter of the
stellar system is comparable to its distance.  This excludes
application of similar models for distant objects in the near
future. A non-parametric method for the deprojection has been recently
developed by Binney \& Gerhard (1996). Its main advantage over
parametric methods is, that it allows to recover more structure
details and the precise radial mass distribution. But an application
of this method to the DIRBE data yielded controversial results
(Binney, Gerhard \& Spergel 1997; in the following: BGS).  The
recovered radial and vertical structure of bar and disk provides a
realistic view of the inner galaxy. But in addition, mass
concentrations on the minor axis have been found in the deprojection,
which seem to indicate that the inner galaxy has significant spiral
structure as well.

In Fig.~\ref{termcmp} we compare the axisymmetric part of the BGS
model (short dashed line) with the axisymmetric model by Kent (1992,
long dashed line) and the $^{12}$CO data from Clemens (1985) and HI
data from Burton \& Liszt (1993). Note that Kents model includes a
dark halo and assumes LSR motion $V_0=234\,\kms$ while the BGS model
was plotted for $V_0=220\,\kms$ and does not include a halo. Increasing
the LSR motion increases the gap between model and observed
terminal curve at the solar circle ($l=90^\circ$), but can be
compensated for with a dark matter component. 

\begin{figure}
\begin{center}
\epsfxsize=11cm \epsfbox{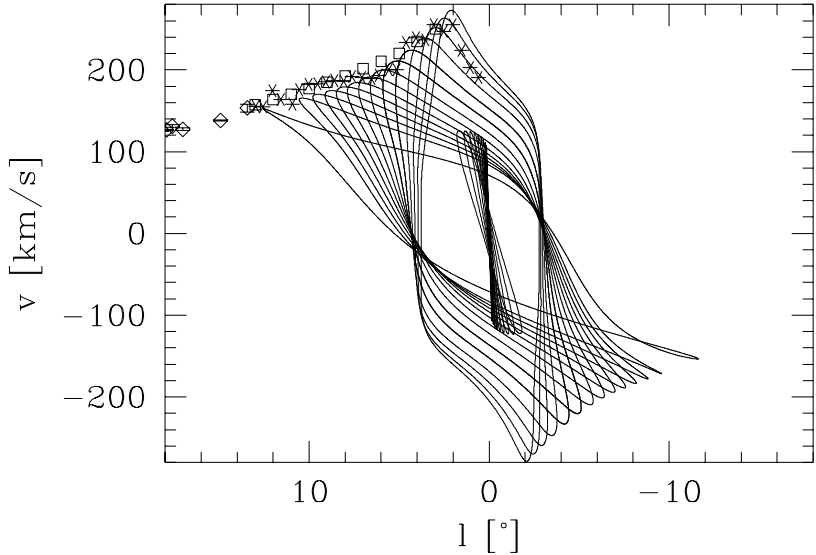}
\vskip -7mm
\end{center}
\caption{The high peak in the observed terminal velocity curve (points) 
is reproduced by the $x_1$-orbital family in the BGS model.}
\label{lvorb}
\end{figure}

Both axisymmetric models fail to explain the peak in the rotation
curve. Between 15 and $40^\circ$ the BGS model fits the data better,
further out the contribution of the dark halo is missing which is
already included in Kents model. When we assume a constant rotation
curve beyond $5\,\kpc$ ($=40^\circ$), the BGS model also matches the
data between $40^\circ$ and $90^\circ$. Closed orbits in the full BGS
potential match the high peak in the rotation curve
(Fig.~\ref{lvorb}).  This confirms that the deprojection indeed
provides a good description of the bar potential. Further out, the
combination of bar and mass concentration on the minor axis
complicates the picture. A hydrodynamical model of the gas flow in the
BGS model is presented below. When the terminal curve for this model
is compared to the data (Fig.~\ref{termcmp}; solid line), the result
depends on the resolution in the gas model. Hydrodynamical forces tend
to depopulate the orbits responsible for the peak in the rotation
curve due to low resolution. A model with higher resolution
matches the peak better (Fig.~\ref{termcmp}; dotted line).

Whether bars and spirals are independent dynamical entities is not yet
clear and may vary from galaxy to galaxy.  Numerical $N$-body
simulations show that a bar can coexist with a spiral of much lower
pattern speed (Sellwood \& Sparke 1988).  On the other hand, detailed
modelling of the observed gas dynamics in NGC 1365 was possible using
bar and spiral perturbation with a single pattern speed (Lindblad,
Lindblad \& Athanassoula 1996). 

Moreover, the spiral mode is most likely growing and decaying within a
few rotations of the galaxy (Lin \& Shu 1964).  For our gas model, we
assumed that these additional light concentrations can be used as a
first approximation to the contribution of spiral arms in this region
and that the spiral pattern is rotating with the same speed as the bar
and is stationary.

A more advanced deprojection technique combined with spiral arm
modeling will be used to shed more light on this issue shortly
(Bissantz \& Gerhard 2000).  This will also allow us to study models
with independently rotating bar and spiral mode.

\section{Steady state / Stationary gas flow solution}

If the gas is allowed to relax in the potential of the galaxy,
i.e. the mass distribution in the galaxy does not change over a few
rotational periods, we expect the gas to form a stationary flow
pattern. Since the gas looses energy in collisions, it arranges itself
to follow nested closed orbits which have no intersections. In barred
galaxies, orbits are organized in families.  The so-called
$x_1$-orbits are elongated along the bar and exist from somewhat
inside the corotation annulus to the center. Inside the bar region
there is a special cusped $x_1$-orbit, which is the last $x_1$ orbit
the gas can settle on. Inside this orbit, the $x_1$ orbits in our
potential develop loops with self-intersections. Gas on such orbits,
would suffer collisions causing it to leave the orbit. In linear
theory, the gas has to pass through the inner Lindblad resonance (ILR), which
causes a phase shift of $\pi/2$ in the orbits. The gas follows this
phase shift with a delayed response, and switches to the $x_2$ family
of orbits, which are elongated perpendicular to the bar and exist only
inside the ILR. In general, further ILRs may exist, but there is no evidence
for further ILRs in the BGS potential. The transition at the ILR is not 
smoothly, but is accompanied by a shock in the gas flow. In strong bars, the
shocks are straight lines, which in real galaxies have been identified
with dust lanes and velocity jumps.  Due to the non-axisymmetric
distribution of gas caused by the shocks or spiral arms, the bar
imposes a gravitational torque on the gas, which causes the gas to
flow in (out) when the trailing spiral arm is inside (outside)
corotation. For leading spiral arms, the torque would be reversed. For
this reason, gas is slowly depleted in the corotation region and
accumulates within the ILR to form a ring or disk of gas on $x_2$ orbits.

\begin{figure}[!t]
\begin{center}
\epsfxsize=\hsize \epsfbox{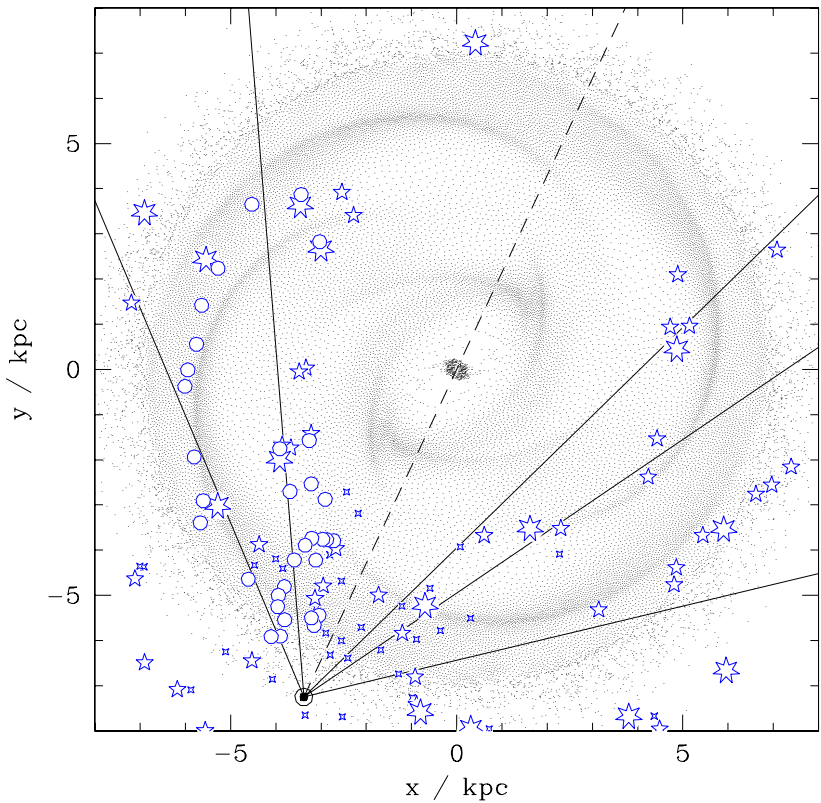}
\end{center}
\vskip -7mm
\caption{Comparison of our model with observed spiral arm tangents (lines,
averaged values) and HII regions (stars) and giant molecular clouds
from Dame \etal. (1986; circles). Rotation is clock wise, and the model is
viewed from the north galactic pole. The bar inclination is $20^\circ$ and
corotation at $3.4\,\kpc$. This model also includes a dark halo.
}
\label{model}
\end{figure}

For hydrodynamics we chose the smooth particle hydrodynamics (SPH)
method (Benz 1990; Steinmetz \& M\"uller 1993). In SPH a continuous
gas distribution is approximated by a spatially smeared out particle
distribution. The advantage of this method is, that self-gravity can
easily be included. Further numerical details of this simulation are given in
Englmaier \& Gerhard (1999). In addition to the free parameters from
the deprojection, our gas model needs only three additional parameters:
the corotation radius or equivalently the pattern speed, 
the effective sound speed, and the LSR motion for calculation 
of (lv) diagrams. Our best model is shown in Fig.~\ref{model}.

The corotation radius in our model is tightly constrained by the
observed (lv)-diagram as follows. Since the 3-kpc-arm is clearly in
non-circular rotation, it must be inside the corotation radius. By
comparison, the 4-armed spiral pattern in the molecular ring between 4
and $7\,\kpc$ shows no large deviation from circular motion. This
observation can only be reproduced in our model, if the corotation is
between 3 and $4\,\kpc$.  The angular extend of the 3-kpc-arm is best
reproduced for corotation at $3.4\,\kpc$.

The $M/L$ ratio and the LSR motion are fixed by a fit of the terminal
velocity curve to the model. Actually the circular rotation velocity
for the LSR is kept constant to $200\,\kms$ times a scaling constant
which depends on $M/L$. The final value for the LSR motion is
$208\,\kms$ after the best model fit has been determined. While this
value is lower than the current best estimate for the LSR motion, we
note that the $M/L$ value and other model details do not depend 
strongly on this parameter. The difference in the terminal curve can
be compensated for by a dark halo.

The gas flow model confirms the allowed range for the bar inclination
found by BGS ($25\pm10$). Models with $20^\circ$ or $25^\circ$ yield
a better overall fits than models with $15^\circ$ or $30^\circ$.
A better model for the stellar
spiral arms may improve this situation.

\section{Identification of spiral arms in the (lv)-diagram}

\begin{figure}[!t]
\begin{center}
\epsfxsize=\hsize \epsfbox{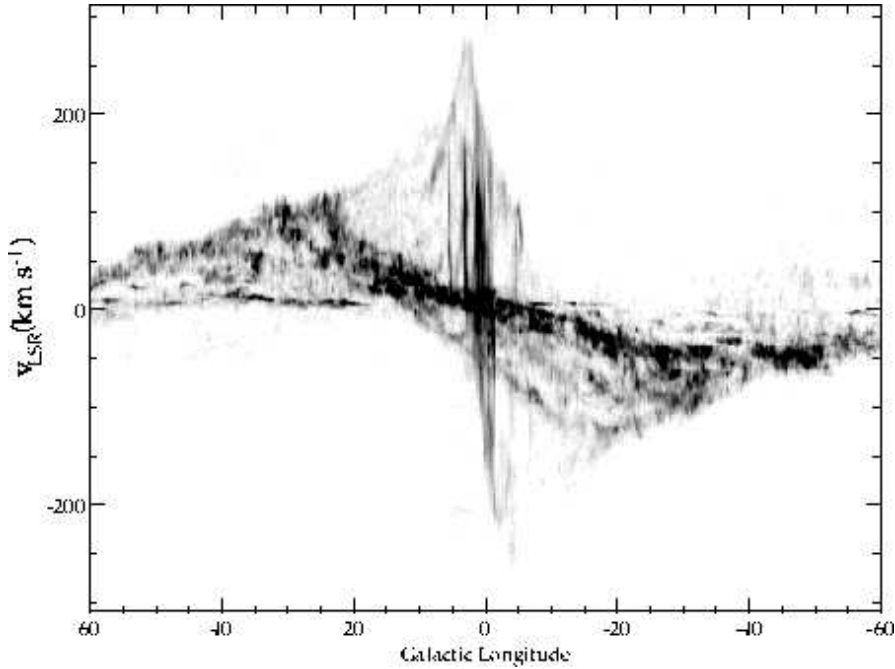}
\end{center}
\vskip -7mm
\caption{ The (lv)-diagram for $^{12}$CO from unpublished data by Dame
\etal. (1998). This figure contains all emission integrated over
latitudes between $b=-2^\circ$ and $b=2^\circ$.  The grey scale is
adjusted such as to emphasize spiral arm structures.  }
\label{co}
\end{figure}

A full sky survey of emission from atomic hydrogen and molecular
species like $^{12}$CO allow mapping of large scale structure such as
spiral arms (Fig.~\ref{co}).  The longitude-velocity diagram
(lv-diagram) shows traces of spiral arms as crowded regions. By
assuming circular rotation, one could map the observed (lv)-position
back to real space, however, this leads to serious errors close to
tangential points (Burton 1971). Nevertheless, with some success, the
principle spiral arm structure has been inferred from
lv-diagrams. First, this was done by Oort, Kerr \& Westerhout (1958),
later the 4-armed structure was first claimed by Georgelin \&
Georgelin (1976) using the (lv)-diagram for large HII-regions.  This
picture has later been refined by Caswell \& Haynes (1987), Lockman
(1989), and others. In Fig.~\ref{model} we compare various known
HII-regions and giant molecular clouds from the literature with our
model.

HII-regions are a particular good tracer for spiral arms, as is known
from observations of external galaxies. The 4 spiral arms are also
prominent in the so-called molecular ring between 4 and $7\,\kpc$.
This area corresponds to a bright band of $^{12}$CO emission between
about $30..60^\circ$ on the northern side of the galaxy (left in Fig.~\ref{co})
to $-(30..60)^\circ$ on the southern side. As Solomon
\etal. (1987) pointed out, the spiral arm tangent at around
$+30^\circ$ is actually split into two components at $25^\circ$ and
$30^\circ$.  A summary of available spiral arm tangent data is listed
in Table~1.

\begin{table}[!t]
\begin{center}
\begin{tabular}{c|ccccc}
Indicator & Scutum    & Aquilla & Centaurus & Norma &  3-kpc  \\
          & far 3-kpc &         &           &       &   \\
\hline
(1) & 29         & 50 & -50  & -32  &     \\
(2) & 24, 30.5 & 49.5 & -50 & -30 &     \\
(3) & 25, 32 & 51  &     &     &     \\
(4) & 25, 30      & 49  &     &     &     \\
(5) & 24, 30      & 47 & -55 & -28 &     \\
(6) & 32         & 46 & -50 & -35 &     \\
(7) & 32      & 48  &-50,-58& -32 & -21 \\
(8) & 29         &    &     & -28 & -21 \\
(9) & 26         &    & -47 & -31 & -20 \\
\hline
\bf average& \bf 30   & \bf 49  & \bf -51 & \bf -31 & \bf -21 \\[1.5em]
Model & & & & &\\
\hline
(a) & $\sim25$& 54& -44 & -33 & -20 \\
(b) & $\sim30$& 50& -46 & -33 & -20 \\
(c) & $\sim29$& 51& -47 & -34 & -22 \\
\end{tabular}

\caption{
Measured spiral arm tangents in inner galaxy:
(1) atomic hydrogen (Weaver 1970; Burton \& Shane 1970; Henderson 1977); 
(2) integrated $^{12}$CO (Cohen \etal. 1980; Grabelsky \etal. 1987);
(3) $^{12}$CO clouds (Dame \etal. 1986);
(4) warm CO clouds (Solomon \etal. 1985);
(5) HII-Regions using H109-$\alpha$ (Lockman 1979; Downes \etal. 1980);
(6) $^{26}$Al from massive stars (Chen \etal. 1996);
(7) Radio $408\,\mhz$ (Beuermann \etal. 1985);
(8) $2.4\,\mym$ (Hayakawa \etal. 1981);
(9) $60\,\mym$ (Bloemen \etal. 1990).
Compared to some models: 
(a) $R_c=3.4\,\kpc$, bar inclination $\varphi=20^\circ$, without halo;
(b) $R_c=3.4\,\kpc$, $\varphi=20^\circ$, with halo $v_0=200\,\kms$;
(c) $R_c=3.4\,\kpc$, $\varphi=25^\circ$, with halo $v_0=200\,\kms$.
}
\end{center}
\end{table}

The molecular emission also shows an arm apparently not traced by
HII-regions: the 3-kpc-arm. This arm shows large non-circular motion,
as it passes in front of the galactic center (where it appears in absorption)
with $+54\,\kms$ radial velocity. All other arms show much less radial velocity
in front of the galactic center.

Our gas model qualitatively reproduces many of the prominent features
in the observed (lv)-diagram (Fig.~\ref{co}).  The 4 spiral arms are
found to be embedded in the molecular ring with about the right
tangential directions (see Table~1).  Our model, however, assumes
perfect point symmetry in the gas, while the galaxy is not. Hence a
perfect match cannot be expected in such an idealized model. Inside
the molecular ring, gas is depleted in the corotation region and the
spiral arms show a gap (see Fig.~\ref{model}). Within corotation the
spiral pattern continues followed by the usual gas flow configuration
in a barred galaxy (see previous section). One of the spiral arms is
similar to the 3-kpc-arm in the galaxy, alas with too small an expansion
velocity.  In this region, the gas flow may be disturbed by the
stellar spiral arms.  In a modified model where we approximate the
spiral arm gravity by the gaseous model spiral arms, but spatially
smeared out to account for wider stellar spiral arms, the 3-kpc-arm
can be reproduced quantitatively (Englmaier \& Gerhard 1999). Our
model also predicts the location of the arm corresponding to the
3-kpc-arm on the far side of the galaxy. It happens to fall close to
the molecular ring and indeed this arm tangent was found to be split
into two components by Solomon \etal. (1987) and is also visible at
$l=25^\circ, 30^\circ$ in Fig.~\ref{co}. A similar position for the
far 3-kpc-arm was suggested by Sevenster (1999), but see Fux (1999a)
for an alternative explanation. A nuclear disk with $\sim 200\,\pc$ radius
is formed in the model by gas on $x_2$-orbits. Size and rotational velocity
of this disk depends on the enclosed mass which has been adjusted by the
central density cusp profile. While the disk matches the observed
CS emission (Stark \etal. 1991), only part of the disk is occupied by
dense enough gas to show CS emission.

\section{OH/IR stars as a fossil dynamical record}

\begin{figure}[p!]
\begin{center}
\epsfxsize=9cm \epsfbox{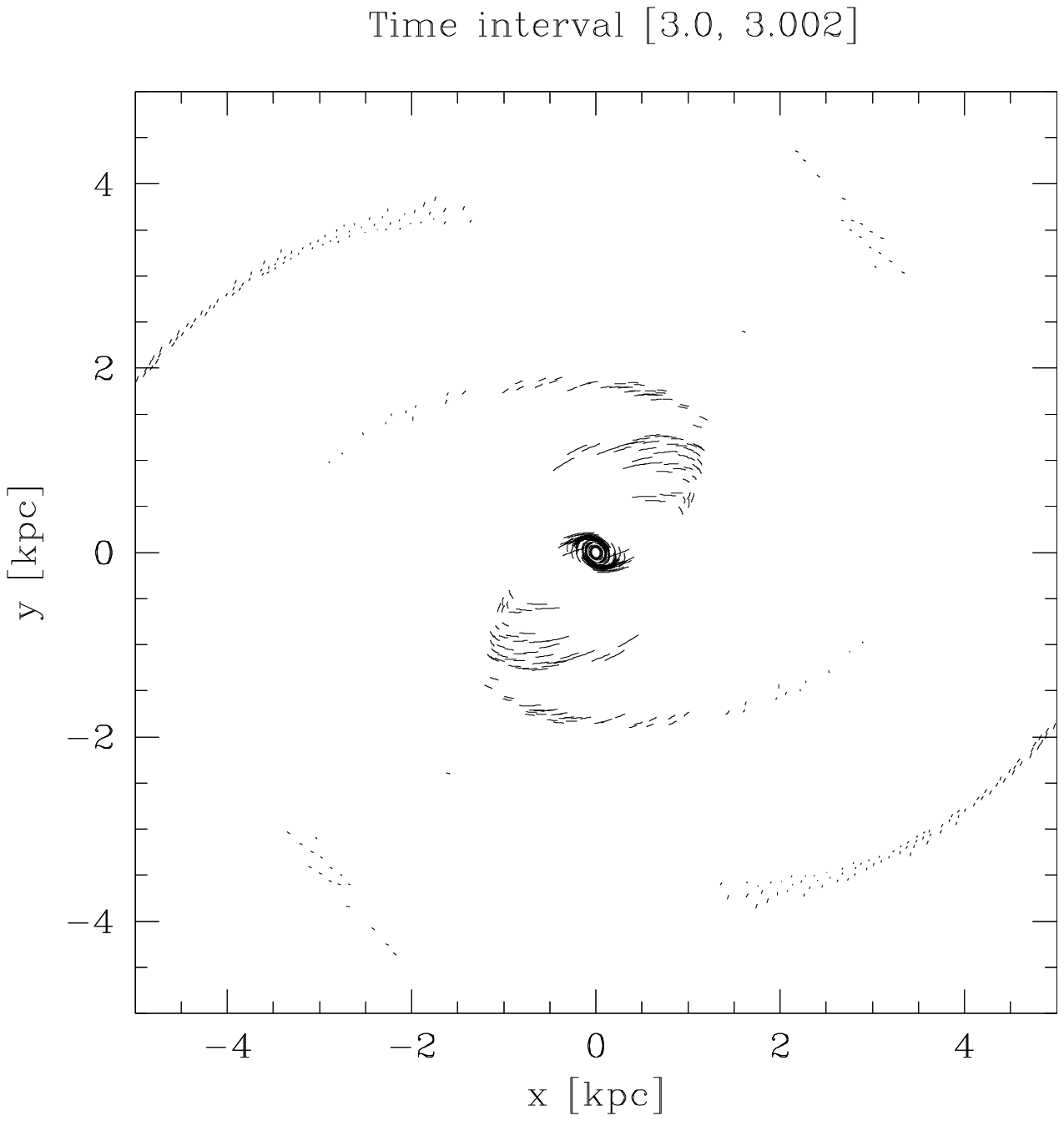}
\vskip -2mm
\epsfxsize=9cm \epsfbox{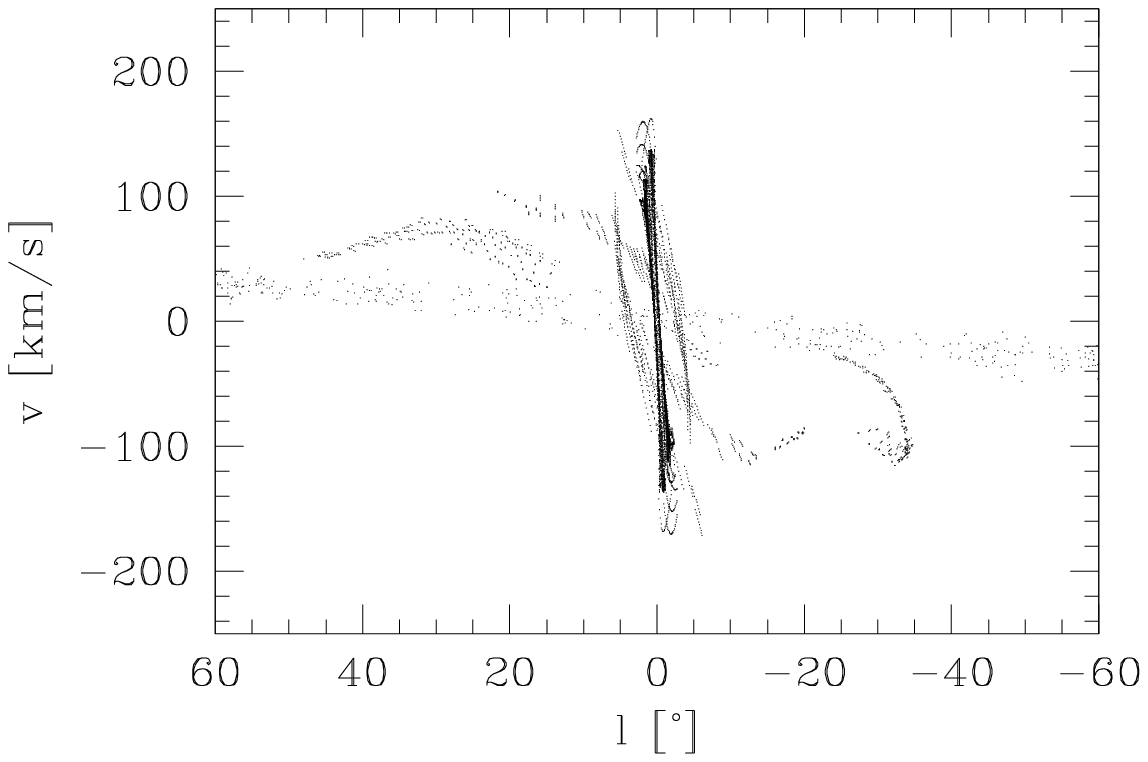}
\vskip -7mm
\end{center}
\caption{Location of star forming regions of compressed gas in face-on
view (upper) and seen in the (lv)-diagram (lower). For each particle a
$2\,\myr$ long trace of its orbit is shown, which is approximately the
duration of the OH/IR phase. The bar is aligned with the diagonal line
from the lower left to the upper right (compare with
figure~\ref{model}). The sun to galactic center line is inclined
$25^\circ$ to the bar. The location of the sun is just outside the
diagram at $x\approx-2.7\,\kpc$, $y\approx-7.5\,\kpc$. The 3-kpc-arm
is the horizontal trace of stars starting at the lower left end of the
bar. For the (lv)-diagram a circular velocity of $220\,\kms$ has been used
for the transformation in the LSR. The 3-kpc-arm passes the galactic
center ($l=0^\circ$) with the velocity $\sim-40\,\kms$ and continues
to $l=-21^\circ$, $v=-100\,\kms$.
}
\label{ohir1}
\end{figure}

Sevenster (1997, 1999) made a complete sample of the OH/IR stars in
the bulge region between $-45^{\circ}\leq l\leq 10^{\circ}$ and
$|b|\leq 3^{\circ}$. The OH/IR stars are giant stars that lose matter
in the so-called asymptotic giant branch (AGB) superwind phase. While
these stars are quite old, i.e. $\sim1$ to $8\,\gyr$, they quickly go
through an evolutionary phase characterized by OH maser emission in
the IR. The maser emission lasts only for $\sim 10^{5-6}\,\yr$ which
is short compared to dynamical timescales.  The sample of OH/IR
therefore provides a snapshot of the dynamically evolved star
formation regions $\sim1$ to $8\,\gyr$ ago. Sevenster (1999) found,
that the lv-diagram for the sample shows a striking correlation
between the OH/IR stars and the 3-kpc-arm observed in $^{12}$CO. A
young group of about 100 to $350\,\myr$ old OH/IR stars follows this
arm very nicely, while the older OH/IR population is more equally
distributed. This appears to be in contradiction with dynamics, since
a circular orbit at $3\,\kpc$ takes $\sim2\pi r/v\sim80\,\myr$ to
complete. Any trace of the spatial distribution of star formation
would have been wiped out. However, if these stars were formed in one
of the spiral arms which itself rotates, then only the relative orbit
between stars and arm would matter.  For a constant pattern speed
equal to the bar pattern speed, say $\Omega\sim60\,\kmskpc$, the orbit in
the rotating frame of the arm would take about $\sim2\pi r/(v-\Omega
r)\sim0.5\,\gyr$ to complete.  This much longer timescale may allow
stars formed at the same place to generate a distribution similar to
the observed one. Furthermore, the 3-kpc-arm is not a strong shock,
making it possible for stars formed within to drift away rather slowly.

\begin{figure}[p!]
\begin{center}
\epsfxsize=9cm \epsfbox{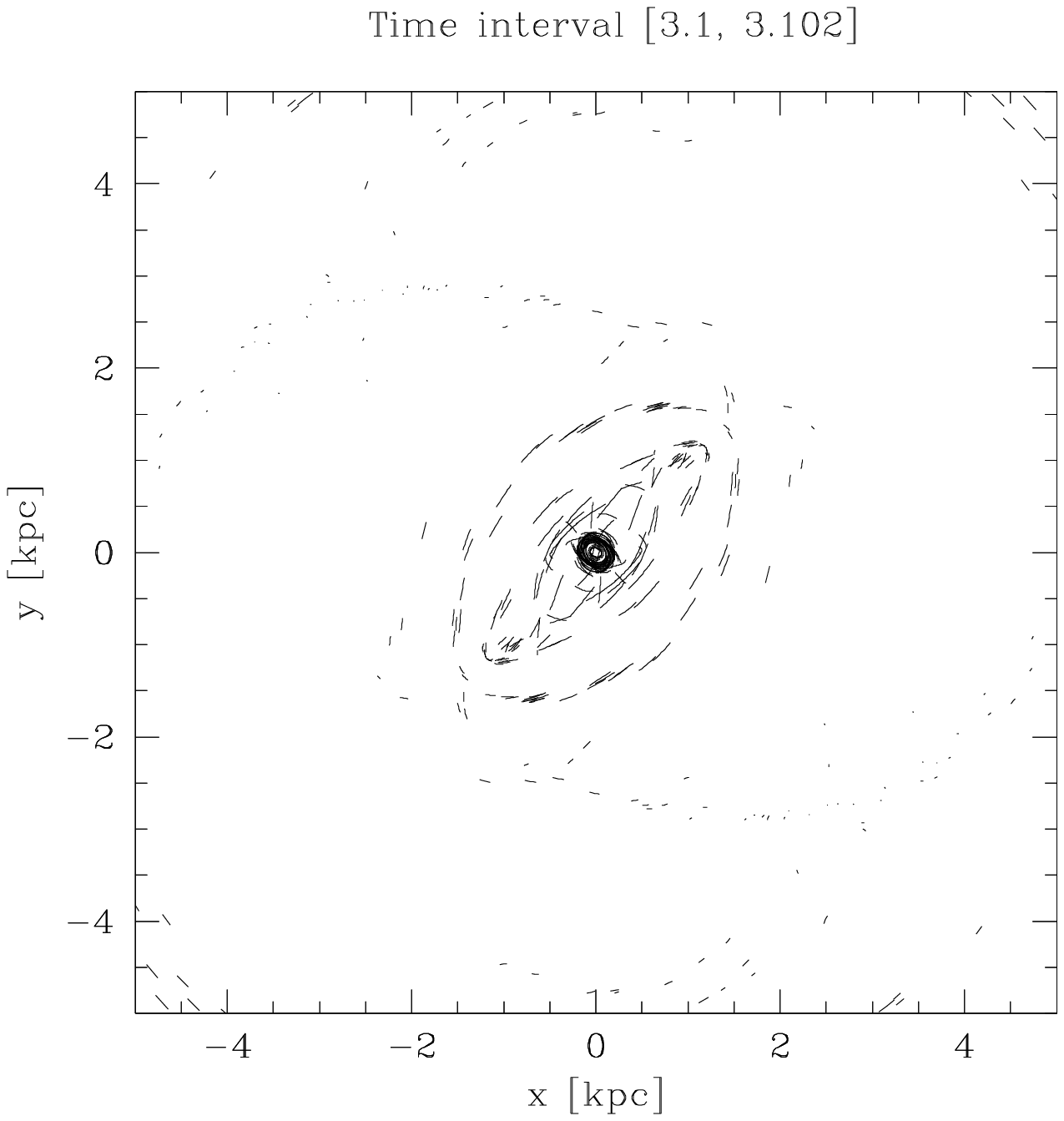}
\vskip -2mm
\epsfxsize=9cm \epsfbox{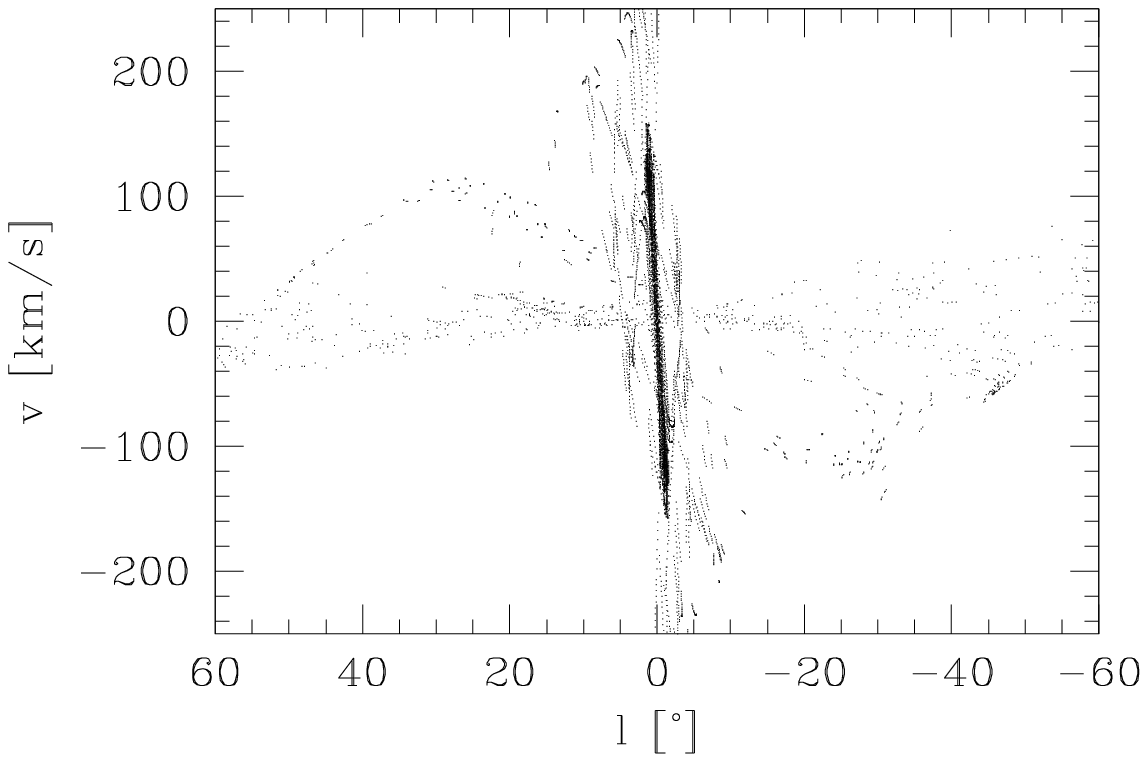}
\vskip -7mm
\end{center}
\caption{Same as figure~\ref{ohir1}, but $100\,\myr$ later. Stars
created in the 3-kpc-arm have formed elliptical rings aligned with
the bar which do not follow the 3-kpc-arm in the (lv)-diagram, but
are restricted to $|l|<10^\circ$. 
However, a trace of stars originally created in the corotation 
region forms a bridge between corotation and ends of the bar 
maintaining a 3-kpc-arm alike structure in the (lv)-diagram.
}
\label{ohir2}
\end{figure}

In order to estimate the dynamical constraints of our model on the
observed OH/IR star distribution, we picked our best model and
selected gas particles which are in compressed areas, i.e. in spiral
arm shocks (see Fig.~\ref{ohir1}).  
We evolved these particles as test particles in the
background potential of the model, and plotted snapshots for the
particle distribution at later times. It turns out that, although OH/IR stars
may form a transient arm like feature, it will only last for
a few $100\,\myr$. In Fig.~\ref{ohir2}, we show the lv-diagram for $100\,\myr$
old stars. There is a 3-kpc-arm-like distribution of stars, but these
stars come from the corotation region at the minor axis, not the
3-kpc-arm.  Just inside corotation, a star formed from the gas moves
faster than the bar pattern, and this gives raise to a coriolis force
pulling the star inwards. A group of freshly formed but isolated stars
would form a stream which may be coincident with the 3-kpc-arm. As we
discussed above, the 3-kpc-arm does not contain HII-regions and
presumably, it does not form stars because it is dynamically hot. Stars
which are formed in the 3-kpc-arm are quickly distributed on closed orbits,
it is therefore possible that OH/IR stars are indeed distributed in
rings indicating radii of more compressed gas in our model, and these
orbits, which are not circular orbits, may be indirectly correlated
with the 3-kpc-arm, because arms which are driven by a bar, often
start with a local density enhancement and strong star formation on
the bar major axis. However, we would expect those rings to be
inclined with respect to the 3-kpc-arm, unless the 3-kpc-arm is not
stationary.

A similar result was found by Fux (1999b), where the observed OH/IR
star distribution could be reproduced for about $40\,\myr$ old
stars. The OH/IR stars in his model originate in the far end of
the bar in an area of enhanced density. Contrary to our model, Fux's
model has the additional advantage, that the 3-kpc-arm changes
rapidly and thus the gas is on more ballistic orbits. He did not
show any evidence, however, that the coincidence of stars and gas
can be maintained over multiple rotations as required by the
observations.  While both models are optimized to fit various aspects
of the Milky Way, they both suggest that OH/IR stars are not directly
tracing spiral arms.

We therefore consider it more likely that the stars formed close to
corotation where timescales are comparable to the age of these stars.
It will be interesting if further evidence can be found, to show that OH/IR
stars follow closed orbits and whether these favored orbits have
changed with time.

\section{Discussion}

We presented a coherent description for the structure and gas
dynamics in the Milky Way reaching from small scale $\sim100\,\pc$
to large scale $\sim\,\kpc$. Our model provides a link between
photometric and gas dynamical observations, but does not intend to
be a a self-consistent treatment of formation and evolution of the 
Milky Way. Idealized assumptions about symmetry and in the description
of the ISM have been made to isolate generic from peculiar features.

We find, that the 3-kpc-arm can be explained qualitatively by the
forcing of the bar. The bar may also be responsible for some of the
spiral structure observed in the gas. Strong bars can drive spiral
arms to and somewhat beyond the outer Lindblad resonance (Mulder
1986).  The deprojection of the near-IR bulge light distribution
introduced additional structure on the minor axis of the bar at around
corotation. When this structure is included into the model, four
spiral arms are driven in the gas similar to the observed spiral arm
structure in this region. The pattern is very similar to a model by
Mulder \& Liem (1986), but the location of the sun is different. 
Interestingly, Mulder (1986) mentions that this pattern can be driven by
a strong bar or by a weak bar plus spiral arms beyond corotation.

When the gravity of the spiral arms is included in the model, the
3-kpc-arm can also be explained quantitatively. This finding, and
in addition our inability to find a fitting model for the 3-kpc-arm
in models without the minor axis structure demonstrates, that the
inner galaxy is affected by both bar and spiral arms.

While spiral arms in barred galaxies usually start at the end of the
bar, they do not need to be driven by the bar (Sellwood \& Sparke 1988),
and might be dynamically independent with much lower pattern speed. No
steady state but maybe a periodic steady state can be found in the gas
flow of such a model.  Independent stellar spiral arms further
complicate the gas dynamics to some extent, because they cause
non-linear velocity jumps in the gas and will not just add to the
structure formed by the bar.  From density wave theory (Lin \& Shu
1964) we know, that spiral arms are not stationary but form and decay
in a few rotational periods.  With a lower pattern speed, the spiral
pattern would extend to larger radii. A model for the galaxy, where
spiral arms and bar have different pattern speeds has still to be
constructed, yet.

Considering all these objections, it is even more surprising how well
our model explains the spiral structure. Both the bar and the spiral
pattern impose similar constraints on the orientation of the pattern,
i.e. the phase angle.  Nevertheless, this overlap might be a mere
coincidence and a better treatment of bar and spiral mode is underway
to investigate other possibilities (Bissantz \etal., in prep.).

\section*{Acknowledgments}
I want to thank O.~Gerhard for his collaboration on this project.
This work was supported by the Swiss NSF grants 21-40'464.94 and
20-43'218.95.

\vspace{0.7cm}
\noindent
{\large{\bf References}}
{\small

\bref Amaral L.H., L\'epine J.R.D., 1997, MNRAS, 286, 885

\bref Backer D.C., Stramek R.A., 1999, ApJ, 524, 805

\bref Becklin E.E., Neugebauer G., 1968, ApJ, 151, 145

\bref Benz W., 1990, in {\em The Numerical Modelling of Nonlinear
Stellar Pulsations}, ed. Buchler J.R., p. 269, Dordrecht, Kluwer

\bref Bertin G., Lin C.C., 1996, MIT Press, Cambridge, Mass.

\bref Beuermann K., Kanbach G., Berkhuijsen E.M., 1985, A\&A, 153, 17

\bref Binney J.J., Gerhard O.E., Stark A.A., Bally J., Uchida K.I.,
1991, MNRAS, 252, 210

\bref Binney J.J., Gerhard O.E., 1996, MNRAS, 279, 1005

\bref Binney J.J., Gerhard O.E., Spergel D.N., 1997, MNRAS, 288, 365

\bref Binney J.J., Tremaine S., 1987, Princeton University Press

\bref Bissantz N., Gerhard O.E., 2000, in prep.

\bref Bloemen J.B.G.M., Deul E.R., Thaddeus P., 1990, A\&A, 233, 437

\bref Burton W.B., 1971, A\&A, 10, 76

\bref Burton W.B., Shane W.W., 1970, IAU Symposium 38, eds. Becker W., 
Contopoulos G., Reidel, Dordrecht, p. 397

\bref Burton W.B., Liszt H.S., 1993, A\&A, 274, 765

\bref Chen W., Gehrels N., Diehl R., Hartmann D., 1996, A\&AS, 120, 315

\bref Clemens D.P., 1985, ApJ, 295, 422

\bref Caswell J.L., Haynes, R.F., 1987, A\&A, 171, 261

\bref Cohen R.J., Cong H., Dame T.M., Thaddeus P., 1980, ApJ, 239, L53

\bref Dame T.M., Elmegreen B.G., Cohen R.S., Thaddeus P., 1986, ApJ, 305, 892

\bref Dame T., \etal., 1987, ApJ, 322, 706

\bref Dehnen W., Binney J.J., 1998, MNRAS, 294, 429

\bref Downes D., Wilson T.L., Bieging J., Wink J., 1980, A\&AS, 40, 379

\bref Dwek E., \etal, 1995, ApJ, 445, 716

\bref Englmaier P., Gerhard O.E., 1999, MNRAS, 304, 512

\bref Feast M., Whitelook P., 1997, MNRAS, 291, 683

\bref Freudenreich H.T., 1998, ApJ, 492, 495

\bref Friedli D., 1999, in {\em The Evolution of Galaxies on
Cosmological Timescales}, eds. Beckman J.E.,  Mahoney T., ASP
Conf. Series, astro-ph/9903143

\bref Fuchs B., M\"ollenhof C., Heidt J., 1998, A\&A, 336, 878

\bref Fux R., 1999a, A\&A, 347, 77

\bref Fux R., 1999b, in {\em The Evolution of Galaxies on Cosmological
Timescales}, eds. Beckman J.E., Mahoney T., ASP Conf. Series,
astro-ph/9908091

\bref Georgelin Y.M., Georgelin Y.P., 1976, A\&A, 49, 57

\bref Gerhard O.E., 1996, in: {\em Unsolved Problems of the Milky
        Way}, IAU Symp.~169, eds. Blitz L., Teuben P., p. 79

\bref Gerhard O.E., 1999, in {\em Galaxy Dynamics}, eds. Merritt D.R.,
Valluri M., Sellwood J.A., ASP Conf. Series, p. 307

\bref Gerhard O.E., Vietri M., 1986, MNRAS, 223, 377

\bref Grabelsky D.A., Cohen R.S., Bronfman L., Thaddeus P., 1987, ApJ,
	315, 122

\bref Hayakawa S. \etal, 1981, A\&A, 100, 116

\bref Henderson A.P., 1977, A\&A, 58, 189


\bref Kent S.M., 1992, ApJ, 387, 181

\bref Kent S.M., Dame T.M., Fazio G., 1991, ApJ, 378, 131

\bref Lin C.C., Shu F.H., 1964, ApJ, 140, 646

\bref Lin C.C., Yuan C., Shu F.H., 1969, ApJ, 155, 721

\bref Lindblad P.A.B., Lindblad P.O., Athanassoula E., 1996, A\&A, 313, 65

\bref Lockman F.J., 1979, ApJ, 232, 761

\bref Lockman F.J., 1989, ApJ Suppl., 71, 469

\bref Mulder W.A., 1986, A\&A, 156, 354

\bref Mulder W.A., Liem B.T., 1986, A\&A, 157, 148

\bref Oort J.H., Kerr F.T., Westerhout G., 1958, MNRAS, 118, 379

\bref Rohlfs K., Kreitschmann J., 1987, A\&A, 178, 95


\bref Schmidt M., 1965, Bull. Astr. Inst. Netherlands, 13, 15

\bref Sevenster M., 1997, thesis, Leiden

\bref Sevenster M., 1999, MNRAS, in press, astro-ph/9907319

\bref Sellwood J.A., Sparke L.S., 1988, MNRAS, 231, 25

\bref Sofue, 1973, PASJ, 25, 207

\bref Solomon P.M., Sanders D.B., Rivolo A.R., 1985, ApJ, 292, 19

\bref Solomon P.M., Rivolo A.R., Barrett J., Yahil A., 1987, ApJ, 319, 730

\bref Stark A.A., Bally J., Gerhard O.E., Binney J.J., 1991, MNRAS, 248, 14

\bref Steinmetz M., M\"uller E., 1993, A\&A, 268, 391

\bref Toomre A., 1977, ARA\&A, 15, 437

\bref Vall\'ee J.P., 1995, ApJ, 454, 119

\bref Wada K., Taniguchi Y., Habe A., Hasegawa T., 1994, ApJ, 437L, 123

\bref Weaver 1970, in: {\em The Spiral Structure of Our Galaxy},
	IAU Symp.~38, eds. Becker W., Contopoulos G., Reidel, Dordrecht, p. 126

\bref Weiner B.J., Sellwood J.A., 1999, ApJ, 524, 112

\bref Wielen R., 1974, PASP, 86, 341

}

\vfill

\end{document}